\documentclass[a4paper]{article}      
\usepackage{amssymb,amsfonts} 
\usepackage{graphics}                 
\usepackage{color}                    
\usepackage{hyperref}                 
\newtheorem{theorem}{Theorem}[section]

\newtheorem{example}{Example}[section]

\newcommand{\qed}{\hfill\rule{2mm}{2mm}}  

\title{\bf Symbolic computation of moments of sampling distributions}
\author{E. Di Nardo \footnote{Dipartimento di Matematica,
 Universit\`a degli Studi della Basilicata
 C.da Macchia Romana, I-85100 Potenza, E-mail: \tt elvira.dinardo@unibas.it},
G. Guarino \footnote{Medical School, Universit\`a Cattolica del Sacro Cuore
(Rome branch), Largo Agostino Gemelli 8, I-00168, Roma, Italy., E-mail: \tt giuseppe.guarino@rete.basilicata.it},
D. Senato \footnote{Dipartimento di Matematica, Universit\`a degli Studi della Basilicata
 C.da Macchia Romana, I-85100 Potenza, E-mail: \tt domenico.senato@unibas.it} }

\begin{document}
\setlength{\baselineskip}{15pt}
\maketitle
\begin{abstract}
By means of the notion of umbrae indexed by multisets, a general
method to express estimators and their products in terms of power
sums is derived. A connection between the notion of multiset and
integer partition leads immediately to a way to speed up the
procedures. Comparisons of computational times with known
procedures show how this approach turns out to be more efficient
in eliminating much unnecessary computation.
\par \medskip \noindent
{\bf keywords: } Umbral calculus; symmetric functions; moments of moments; sam\-pling distributions; $U$-statistics
\par \medskip \noindent
\end{abstract}
\section{Introduction}

It is acknowledged that an appropriate choice of language and
notation can simplify and clarify many statistical calculations.
In recent years, most of the work has been done by symbolic
computation, main references are \cite{McCullagh1},
\cite{Andrews2}. These books offer a variety of applications of
symbolic methods, from asymptotic expansions to the Edgeworth
series, from likelihood functions to the saddlepoint
approximations. In \cite{Zeilrberg}, Zeilberger describes a methodology for using
computer algebra systems to automatically derive moments, up to
order $4,$ of interesting combinatorial random variables. Such a methodology is
applied to pattern statistics of permutations. For
different applications of computer algebra in statistics, see also
\cite{Wynn}.

The aim of this paper is to show the computational efficiency of
umbral calculus in manipulating expressions involving random
variables.

The umbral calculus to which we refer is the version featured by
Rota and Taylor in \cite{SIAM}. The basic device is the representation of a unital
sequence of numbers by a symbol $\alpha,$ called an {\it umbra},
that is, the sequence $1,a_1,a_2, \ldots$ is represented by the
sequence $1,\alpha,\alpha^2,\ldots$ of powers of $\alpha$ via an
operator $E$ resembling the expectation operator of random
variables. This approach has led to a finely adapted language for
random variables by Di Nardo and Senato, see \cite{DiNardo}. In \cite{DiNardo1},
attention is focused on cumulants since a random variable is often
better described by its cumulants than by its moments, as it
happens for the family of Poisson random variables. Moreover, due
to the properties of additivity and invariance under translation,
cumulants are not necessarily connected with moments of any
probability distribution.  As a matter of fact, an umbra seems to
have the structure of a random variable but with no reference to a
probability space, bringing us closer to statistical methods. In
\cite{DiNardo3}, it is shown that classical umbral calculus
provides a unifying framework for unbiased estimators of
cumulants, called $k$-statistics, and their multivariate
generalizations. Moreover, within the umbral framework, a
statistical result does not require a check of background details
by hand, but becomes a corollary of a more general theorem.

Here, we focus attention on the more general problem of
calculations of algebraic expressions such as the variance of a
sample mean or, more generally, moments of sampling distributions,
which have a variety of applications within statistical inference
\cite{StuartOrd}. The field has, in the past, been marred by the
difficulty of manual computations. Symbolic computations have
removed many of such difficulties, leaving some issues unresolved.
One of the most intriguing questions is to explain why symbolic
procedures, which are straightforward in the multivariate case,
turn out to be obscure in the simpler univariate one, see
\cite{Andrews2}.

We show that the notion of multiset is the key for dealing with
symbolic computation in multivariate statistics. Actually, at the
root of the question, there are some aspects of the combinatorics
of symmetric functions that would benefit if the attention is
shifted from sets to the more general notion of multiset. On the
other hand, due to its generality, umbral calculus reduces the
combinatorics of symmetric functions, commonly used by
statisticians, to few relations which cover a great variety of
calculations. In particular umbral equivalences (\ref{(ter3)})
smooth the way to handle any kind of product of sums. As example,
by using equivalences (\ref{(ter3)}), we evaluate the mean of
product of augmented polynomials in separately independent and
identically distributed random variables. The resulting umbral
strategy is completely different from those recently proposed in
the literature, see for instance \cite{Vrbik}, and computationally
more efficient, as we show in the last section.

Moreover, by means of the notion of umbrae indexed by multiset, we
remove the necessity to specify if the random variables of a
vector are identically distributed or not. The basic procedure
consists in finding multiset subdivisions, which suitably extends
the notion of set partitions. The strategy here proposed is a sort
of iterated inclusion-exclusion rule \cite{Andrews1, Andrews}, whose
efficiency is improved taking into account the structure of
multiset and its relation with integer partitions. The result is
the algorithm {\tt makeTab}, given in the appendix.

The paper is structured as follows. Section 2 is provided for
readers unaware of classical umbral calculus. We resume
terminology, notation and some basic definitions. In Section 3, we
give a general procedure for writing down $U$-statistics. Recall
that many statistics of interest may be exactly represented or
approximated by $U$-statistics \cite{Hoeffding}. Such a procedure
is based on the umbral relation between moments and augmented
symmetric functions. The connection with power sums is analyzed in
Section 4. The effectiveness of umbral methods is shown in several
examples, proposed with the intention of helping the reader
unaware of umbral calculus to understand the basic
algebraic rules necessary to work with this syntax. Section 5 is
devoted to umbral formulae giving power sums in terms of augmented
symmetric functions. These formulae provide the most natural way
to form the product of augmented symmetric functions by using a
suitable umbral substitution. The consequent reduction of the
computational time is made clear through some examples, which
point out the role played by the singleton umbra in selecting the
suitable variables. Section 6 is devoted to computational
comparisons with the procedures known in the literature in dealing
with moments of sampling distributions. The speed up of umbral
methods is evident. Some concluding remarks end the paper.

Although, by a numerical point of view, {\tt MAPLE} seems to work
less efficiently respect to {\tt MATHEMATICA}, we have implemented
our algorithms in {\tt MAPLE} because the syntax is more
comfortable for symbolic computation. All tasks have been
performed on a PC Pentium(R)4 Intel(R), CPU 3.00 Ghz, 480MB Ram
with {\tt MAPLE} version 10.0 and {\tt MATHEMATICA} version 4.2.
\section{The classical umbral calculus}
Classical umbral calculus is a syntax consisting of the following data:
\begin{enumerate}
\item[{\it i)}]  a set $A=\{\alpha,\beta, \ldots \},$ called the
{\it alphabet}, whose elements are named {\it umbrae}; \item[{\it
ii)}] a commutative integral domain $R$ whose quotient field is of
characteristic zero; \item[{\it iii)}] a linear functional $E,$
called an {\it evaluation}, defined on the polynomial ring $R[A]$
and taking values in $R$ such that
\begin{enumerate}
\item[{\it a)}] $E[1]=1;$ \item[{\it b)}] $E[\alpha^i \beta^j
\cdots \gamma^k] = E[\alpha^i]E[\beta^j] \cdots E[\gamma^k]$ for
any set of distinct umbrae in $A$ and for $i,j,\ldots,k$
non-negative integers ({\it uncorrelation property});
\end{enumerate} \item[{\it iv)}] an element $\varepsilon \in A,$
called an {\it augmentation}, such that $E[\varepsilon^n] = 0$ for
all $n \geq 1;$ \item[{\it v)}] an element $u \in A,$ called a
{\it unity} umbra, such that $E[u^n]=1$ for all $n \geq 1.$
\end{enumerate}

Note that, for statistical applications, $R$ is the field of real numbers.

An {\it umbral polynomial} is a polynomial $p \in R[A].$ The
support of $p$ is the set of all umbrae occurring in $p.$
\par
If $p$ and $q$ are two umbral polynomials, then
\begin{enumerate}
\item[${\it i)}$] $p$ and $q$ are {\it uncorrelated} if and only
if their supports are disjoint; \item[${\it ii)}$] $p$ and $q$ are
{\it umbrally equivalent} iff $E[p]=E[q],$ in symbols $p \simeq
q.$
\end{enumerate}
The basic idea of the classical umbral calculus is to associate a
sequence of numbers $1, a_2, a_3, \ldots$ to an indeterminate
$\alpha,$ which is said to represent the sequence. This device is
familiar in statistics, when $a_i$ represents the $i$-th moment of
a random variable $X.$ In this case, the sequence $1, a_1, a_2,
\ldots$ results from applying the expectation operator $E$ to the
sequence $1, X, X^2, \ldots$ consisting of powers of $X.$ This is
why the elements $a_n \in R$ such that
$$E[\alpha^n]=a_n, \,\, n \geq 0$$
are named {\it moments} of the umbra $\alpha$ and we say that the
umbra $\alpha$ {\it represents} the sequence of moments
$1,a_1,a_2,\ldots.$ The umbra $\epsilon$ plays the same role of a
random variable which takes the value $0$ with probability 1 and
the umbra $u$ plays the same role of a random variable which takes
the value $1$ with probability 1. The uncorrelation property among
umbrae parallels the analogue one for random variables. In this
setting no attention must be paid to the well-known \lq\lq moment
problem\rq\rq.
\par
In parallel with random variable theory, the {\it factorial
moments} of an umbra $\alpha$ are the elements $a_{(n)} \in R$
corresponding to the umbral polynomials $(\alpha)_n = \alpha
(\alpha -1) \cdots (\alpha-n+1), \, n \geq 1,$ via the evaluation
$E,$ that is $E[(\alpha)_n]=a_{(n)}.$

There are umbrae playing a special role in the umbral calculus.
Their properties have been investigated with full particulars in
\cite{DiNardo, DiNardo1}.
\begin{quote} {\bf Singleton umbra.} The {\it singleton umbra} $\chi$ is
the umbra whose moments are all zero, except the first
$E[\chi]=1.$ Its factorial moments are $x_{(n)} =
(-1)^{n-1}(n-1)!$ As we will see later on, this umbra is the
keystone for managing symmetric umbral polynomials.
\end{quote}
\begin{quote} {\bf Bell umbra.} The {\it Bell umbra} $\beta$ is the umbra whose factorial moments are all equal to $1,$ that is
$E[(\beta)_n]=1$ for all $n \geq 1.$ Its moments are the Bell
numbers, that is the number of partitions of a finite nonempty set
with $n$ elements, or the $n$-th coefficient in the Taylor series
expansion of the function $\exp(e^t-1).$ So $\beta$ is the umbral
counterpart of a Poisson random variable with parameter $1.$
\end{quote}
It is possible that two distinct umbrae represent the same
sequence of moments, in such case these are called {\it similar
umbrae}. More formally two umbrae  $\alpha$ and $\gamma$ are said
to be {\it similar} when
$$E[\alpha^n]=E[\gamma^n] \,\,\, \forall \, n \geq
0, \quad \hbox{in symbols} \,\,  \alpha \equiv \gamma.$$
Furthermore, given a sequence $1,a_1,a_2,\ldots$ in $R,$ there are
infinitely many distinct, and thus similar umbrae representing the
sequence. So, the umbral counterpart of a univariate random sample
is a $n$-vector $(\alpha_1,\alpha_2, \ldots, \alpha_n),$ where
$\alpha_i, i =1,2,\ldots,n$ are uncorrelated umbrae, similar to
the same umbra $\alpha.$ Thanks to the notion of similar umbrae,
it is possible to extend the alphabet $A$ with the so-called {\it
auxiliary umbrae} resulting from operations among similar umbrae.
This leads to construct a {\it saturated umbral calculus} in which
auxiliary umbrae are handled as elements of the alphabet
\cite{SIAM}. In the following, we focus the attention on
auxiliary umbrae which play a special role. Let
$\{\alpha_1,\alpha_2,\ldots,\alpha_n\}$ be a set of $n$
uncorrelated umbrae similar to an umbra $\alpha.$ The symbol $n .
\alpha$ denotes an auxiliary umbra similar to the sum
$\alpha_1+\alpha_2+\cdots+\alpha_n.$  So $n.\alpha$ is the umbral
counterpart of a sum of independent and identically distributed
random variables. The symbol $\alpha^{.n}$ is an auxiliary umbra
denoting the product $\alpha_1 \, \alpha_2 \, \cdots \, \alpha_n.$
\par
Moments of $\alpha^{.n}$ can be easily recovered from its
definition. Indeed, if the umbra $\alpha$ represents the sequence
$1, a_1, a_2, \ldots,$ then $E[(\alpha^{.n})^k]=a_k^n$ for
nonnegative integers $k$ and $n.$

Moments of $n.\alpha$ can be expressed through integer partitions.
Recall that a partition of an integer $i$ is a sequence
$\lambda=(\lambda_1,\lambda_2,\ldots,\lambda_t),$ where
$\lambda_j$ are weakly decreasing integers and $\sum_{j=1}^t
\lambda_j = i.$ The integers $\lambda_j$ are named {\it parts} of
$\lambda.$ The {\it length} of $\lambda$ is the number of its
parts and will be indicated by $\nu_{\lambda}.$ A different
notation is $\lambda=(1^{r_1},2^{r_2},\ldots),$ where $r_j$ is the
number of parts of $\lambda$ equal to $j$ and $r_1 + r_2 + \cdots
= \nu_{\lambda}.$ For example, $(1^3, 2^1, 3^2)$  is a partition
of the integer $11.$ We use the classical notation $\lambda \vdash
i$ to denote that \lq\lq $\lambda$ is a partition of $i$\rq\rq. By
using an umbral version of the well-known multinomial expansion
theorem \cite{Comtet}, we have
\begin{equation}
(n.\alpha)^i \simeq \sum_{\lambda \vdash i}(n)_{\nu_{\lambda}}
d_{\lambda} \alpha_{\lambda}, \label{(momdot1)}
\end{equation}
where the sum is over all partitions $\lambda= (1^{r_1}, 2^{r_2}, \ldots)$
of the integer $i,$  $(n)_{\nu_{\lambda}} = 0$ for $\nu_{\lambda} > n,$
\begin{equation} d_{\lambda}  = \frac{i!}{r_1!r_2!\cdots} \,
\frac{1}{(1!)^{r_1}(2!)^{r_2}\cdots} \quad \hbox{and} \quad
\alpha_{\lambda} \equiv (\alpha_{j_1})^{.r_1}
(\alpha_{j_2}^2)^{.r_2} \cdots,  \label{(not)}
\end{equation}
with $\{j_i\}$ distinct integers chosen in $\{1,2,\ldots,n\}=[n].$
\par
The reader interested in proofs of identities involving auxiliary
umbrae is referred to \cite{DiNardo}.

A feature of the classical umbral calculus is the construction of
new auxiliary umbrae by suitable symbolic substitutions.  For
example, in $n.\alpha$ replace the integer $n$ by an umbra
$\gamma.$ From (\ref{(momdot1)}), the new auxiliary umbra
$\gamma.\alpha$ has moments
\begin{equation}
(\gamma.\alpha)^i \simeq \sum_{\lambda \vdash i}
(\gamma)_{\nu_{\lambda}} d_{\lambda} \alpha_{\lambda}
\label{(ombdot1)}
\end{equation}
and it is called {\it dot-product} of $\gamma$ and $\alpha.$ The
auxiliary umbra $\gamma.\alpha$ is the umbral counterpart of a
random sum. In the following, we recall some useful dot-products
of umbrae, whose properties have been investigated with full
particulars in \cite{DiNardo1}.
\begin{quote} {\bf $\alpha$-factorial umbra.} The umbra $\alpha.\chi$ is called the {\it $\alpha$-factorial} umbra.
Its moments are the factorial moments of $\alpha,$ that is
$(\alpha.\chi)^i \simeq (\alpha)_i.$ If $\alpha \equiv \chi,$ then
$E[(\chi.\chi)^i] = E[(\chi)_i] = x_{(i)} = (-1)^{i-1} (i-1)!.$
\end{quote}
\begin{quote} \label{Ex1} {\bf $\alpha$-cumulant umbra.} The umbra $\chi.\alpha,$ with $\chi$ the singleton umbra, is
called the {\it $\alpha$-cumulant} umbra. By virtue of
(\ref{(ombdot1)}), its moments are
\begin{equation}
(\chi.\alpha)^i \simeq \sum_{\lambda \vdash i} x_{(\nu_{\lambda})}
\, d_{\lambda} \, \alpha_{\lambda} \simeq  \sum_{\lambda \vdash i}
(-1)^{\nu_{\lambda}-1} (\nu_{\lambda}-1)! \, d_{\lambda} \, \alpha_{\lambda}. \label{(cum)}
\end{equation}
Since the second equivalence in (\ref{(cum)}) recalls the
well-known expression of cumulants in terms of moments of a random
variable, it is straightforward to refer the moments of the
$\alpha$-cumulant umbra $\chi.\alpha$ as cumulants of the umbra
$\alpha.$
\end{quote}
\section{$U$-statistics}
In the following, we focus our attention on two kinds of auxiliary
umbrae: $n.\alpha$ and $n.(\chi \alpha).$ Such umbrae, and their
products, are similar to some well-known symmetric polynomials.
Indeed, by definition we have
$$n.\alpha^r \equiv \alpha_1^r + \cdots + \alpha_n^r,$$
where $\alpha_1,\alpha_2,\ldots,\alpha_n$ are uncorrelated umbrae,
similar to the umbra $\alpha.$ Since the umbrae $\alpha_i$ for
$i=1,2,\ldots,n$ can be rearranged without effecting the
evaluation $E,$ the auxiliary umbra $n.\alpha^r$ is similar to the
$r$-th power sum symmetric polynomial in the indeterminates
$\alpha_1,\alpha_2,\ldots,\alpha_n.$
\par
Moreover, since
$$n.(\chi \alpha) \equiv \chi_1 \alpha_1 + \cdots + \chi_n \alpha_n,$$
powers of $n.(\chi \alpha)$ are umbrally equivalent to the umbral
elementary symmetric polynomials $[n.(\chi \alpha)]^k \simeq k!
e_k(\alpha_1,\alpha_2,\ldots,\alpha_n),$ where
$$e_k(\alpha_1,\alpha_2,\ldots,\alpha_n) = \sum_{1 \leq j_1<j_2<\cdots<j_k \leq n} \alpha_{j_1} \alpha_{j_2}
\cdots \alpha_{j_k}, \quad k=1,2,\ldots,n.$$ The proof is given in
\cite{DiNardo3} and it relies on the role played by the umbra
$\chi$ in picking out the indeterminates.
\begin{example}
{\rm If $k=2,$ then
\begin{eqnarray*}
[n.(\chi \alpha)]^2 & \simeq &  (\chi_1 \alpha_1 + \cdots + \chi_n \alpha_n)^2
 \simeq  \sum_{i=1}^n (\chi_i \alpha_i)^2 + 2 \sum_{1 \leq j_1 < j_2 \leq n} \chi_{j_1} \alpha_{j_1}
\chi_{j_2} \alpha_{j_2} \\
& \simeq & 2 \sum_{1 \leq j_1 < j_2 \leq n} \alpha_{j_1} \alpha_{j_2}.
\end{eqnarray*}
The last equivalence follows by observing that $(\chi_i
\alpha_i)^2 \simeq 0$ for $i=1,2,\ldots,n$ since $E[(\chi_i
\alpha_i)^2] = E[\chi_i^2]E[\alpha^2_i]$ for the uncorrelation
property between $\alpha_i$ and $\chi_i,$ and
$E[\chi_i^2]E[\alpha^2_i] =0$ because $E[\chi_i^2]=0.$ On the
other hand, we have $\chi_{j_1} \alpha_{j_1} \chi_{j_2}
\alpha_{j_2} \simeq \alpha_{j_1} \alpha_{j_2}$ for $1 \leq j_1<j_2
\leq n$ since the uncor\-relation pro\-perty among $\chi_{j_1},
\alpha_{j_1}, \chi_{j_2}, \alpha_{j_2}$ implies $E[\chi_{j_1}
\alpha_{j_1} \chi_{j_2} \alpha_{j_2}] =
E[\chi_{j_1}]E[\alpha_{j_1}] E[\chi_{j_2}]E[\alpha_{j_2}],$ but
$E[\chi_{j_1}]E[\alpha_{j_1}]E[\chi_{j_2}]E[\alpha_{j_2}] =
E[\alpha_{j_1}] E[\alpha_{j_2}]$ because $E[\chi_{j_1}] =
E[\chi_{j_2}] =1$ $\Box$.}
\end{example}
The auxiliary umbra $n.(\chi \alpha)$ enables us to rewrite umbral
augmented symmetric polynomials in a very compact
expression. Let $\lambda=(1^{r_1}, 2^{r_2},\ldots)$ be a partition
of the integer $i \leq n.$ Augmented monomial symmetric
polynomials in the indeterminates
$\alpha_1,\alpha_2,\ldots,\alpha_n$ are defined as
$$\tilde{m}_{\lambda}(\alpha_1,\alpha_2,\ldots,\alpha_n) = \sum_{j_1 \ne \ldots \ne j_{r_1} \ne
j_{r_1+1} \ne \ldots \ne j_{r_1+r_2} \ne \ldots } \alpha_{j_1}
\cdots \alpha_{j_{r_1}} \alpha^2_{j_{r_1+1}} \cdots
\alpha^2_{j_{r_1+r_2}} \cdots.$$ In statistical literature, a more
common notation is $[1^{r_1} 2^{r_2} \ldots \,]$ \cite{Lukas}.
For instance, $[1^2 \, 3 \,]$ denotes
$$
\sum_{1 \leq j_1 \ne j_{2} \ne j_{3} \leq n} \alpha_{j_1} \alpha_{j_{2}} \alpha^3_{j_3}.
$$
If $\lambda=(1^{r_1}, 2^{r_2},\ldots) \vdash i,$ then
\begin{equation}
[n.(\chi \alpha)]^{r_1} [n.(\chi \alpha^2)]^{r_2} \cdots \simeq
[1^{r_1} 2^{r_2} \ldots \,], \label{(aug)}
\end{equation}
taking into account the role played by the umbra $\chi$ in
selecting variables. We point out that the umbral notation is very
similar to the notation $[1^{r_1} 2^{r_2} \ldots \,].$ As before,
we give an example in order to clarify equivalence (\ref{(aug)}).
\begin{example}
{\rm If $\lambda =(1^2,3),$ then
\begin{eqnarray*}
& & [n.(\chi \alpha)]^{2} [n.(\chi \alpha^3)] \simeq (\chi_1
\alpha_1 + \cdots + \chi_n \alpha_n)^2
(\chi_1 \alpha_1^3 + \cdots + \chi_n \alpha_n^3) \\
& \simeq & \left( \sum_{1 \leq j_1 \ne j_2 \leq n} \chi_{j_1}
\alpha_{j_1} \chi_{j_2} \alpha_{j_2} \right) \left( \sum_{1 \leq
j_3  \leq n} \chi_{j_3} \alpha_{j_3}^3 \right) \simeq \sum_{1 \leq
j_1 \ne j_2 \ne j_3 \leq n} \alpha_{j_1} \alpha_{j_2}
\alpha_{j_3}^3.
\end{eqnarray*}
The last equivalence follows by observing that $E[\chi_{j_1}
\chi_{j_2} \chi_{j_3}]$ vanishes where there is at least one pair
of equal indexes $\Box$.}
\end{example}
In the following theorem, we give the umbral formulation of a
fundamental expectation result in statistics, see \cite{StuartOrd}.
This is a deep result because it lies at the core of unbiased
estimation and moments of moments literature.
\begin{theorem} \label{ttt}
If $\lambda=(1^{r_1},2^{r_2},\ldots)$ is a partition of the
integer $i \leq n,$ then
\begin{equation}
\alpha_{\lambda} \simeq \frac{1}{(n)_{\nu_{\lambda}}} [n.(\chi\alpha)]^{r_1}[n.(\chi\alpha^2)]^{r_2}\cdots.
\label{(ustat)}
\end{equation}
\end{theorem}
See \cite{DiNardo3} for the proof.

Equivalence (\ref{(ustat)}) states how to estimate products of
moments $\alpha_{\lambda}$ by means of only $n$ bits of
information drawn from the population. In umbral terms, the
population is represented by $\alpha$ and the $n$ bits of
information are the uncorrelated umbrae $\alpha_1,
\alpha_2,\ldots, \alpha_n,$ coming into
$[n.(\chi\alpha^i)]^{r_i}.$ Moreover, having shown that the umbral
polynomials $[n.(\chi\alpha^i)]^{r_i}$ are similar to elementary
polynomials, Theorem \ref{ttt} discloses a more general result:
products of moments are umbrally equivalent to products of umbral
elementary polynomials. The symmetric polynomial on the right side
of equivalence (\ref{(ustat)}) is named $U$-{\it statistic} of
uncorrelated and similar umbrae
$\alpha_1,\alpha_2,\ldots,\alpha_n.$ We take a moment to motivate
this denomination. Usually an $U$-statistic has the form
$$U=\frac{1}{(n)_k}\sum \Phi(X_{j_1}, X_{j_2}, \ldots, X_{j_k}),$$
where $X_1, X_2, \ldots, X_n$ are $n$ independent random
variables, and the sum ranges in the set of all permutations
$(j_1,j_2,\ldots,j_k)$ of $k$ integers with $1 \leq j_i \leq n.$
If $X_1,X_2, \ldots, X_n$ have the same cumulative distribution
function $F(x),$ $U$ is an unbiased estimator of the population
parameter
$$\theta(F)=\int \cdots \int \Phi(x_1,\ldots,x_k) dF(x_1)\cdots dF(x_k).$$
In this case, the function $\Phi$ may be assumed to be a symmetric
function of its arguments. Often, in the applications, $\Phi$ is a
polynomial in $X_i$'s so that the $U$-statistic is a symmetric
polynomial. Hence, by virtue of the fundamental theorem on
symmetric polynomials, such an $U$-statistic can be expressed as a
polynomial in elementary symmetric polynomials.
\begin{example}{\it Moment powers.}
{\rm Let us consider the partition $\lambda=(1^2)$ of the integer 2.
The symmetric polynomial
$$U= \frac{1}{(n)_2} [n.(\chi \alpha)]^2, \qquad n \geq 2,$$
is the $U$-statistic related to $\alpha^{.2} \simeq a_1^2.$ Indeed,
setting $r_1=2$ and $\nu_{\lambda}=2$ in (\ref{(ustat)}),
we have
$$\alpha^{.2} \simeq \frac{1}{(n)_2} \sum_{i\ne j} \alpha_i \alpha_j  \simeq U,$$
where the last equivalence follows by expanding the square of
$n.(\chi \alpha).$  $\Box$}
\end{example}
\begin{example}{\it $k$-statistics.} \label{kstat1}
{\rm The $i$-th $k$-statistic $k_i$ is the unique symmetric
unbiased estimator of the cumulant $\kappa_i$ of a given
statistical distribution, that is $E[k_i] = \kappa_i$
\cite{StuartOrd}. In umbral terms, we have
\begin{equation}
(\chi.\alpha)^i \simeq \sum_{\lambda \vdash i}
\frac{x_{\nu_{\lambda}}}{(n)_{\nu_{\lambda}}} \, d_{\lambda}
[n.(\chi\alpha)]^{r_1}[n.(\chi\alpha^2)]^{r_2}\cdots, \label{(kk)}
\label{(kstat)}
\end{equation}
by using equivalence (\ref{(cum)}) and Theorem \ref{ttt}. Equivalence (\ref{(kk)})
is the umbral version of the  $i$-th  $k$-statistic $k_i.$  $\Box$}
\end{example}
Usually $k$-statistics are expressed in terms of power sums in the
data points, $S_r = \sum_{i=1}^n X_i^r.$ Umbrally, this is
equivalent to expressing  $k$-statistics in terms of $n.\alpha^r,$
that is to expressing products of auxiliary umbrae such as
$[n.(\chi \alpha^i)]^{r_i}$ in terms of $n.\alpha^j,$ for some
$j.$ Next section is devoted to exploring such relations, which
also allow us to express moments of sampling distributions in
terms of population moments.
\section{Augmented and power sums symmetric functions}

In this section we turn our attention to symmetric functions
useful in computing moments of sampling distributions, i.e
augmented monomial symmetric functions and power sums, with
special care in formula converting the former in terms of the
latter and viceversa. Such polynomials are classical bases of the
algebra of symmetric polynomials. The well-known changes of bases
involve the lattice of partitions, see \cite{Stanley}. Several
packages are available aiming to implement changes of bases (see
http://garsia.math.yorku.ca/MPWP/). For instance, the {\tt SF}
package \cite{Stembridge} is an integrated  {\tt MAPLE} package
devoted to symmetric functions. The use of such packages requires
a good knowledge of symmetric function theory and is not so
obvious. Moreover, due to their generality, such packages are slow
when applied to large variable sets.

The connection between augmented symmetric functions and power
sums has been given in umbral terms, this because umbral
notation simplifies the changes of bases, taking advantage of
multiset notion. In the following we summarize the steps necessary
to construct such formulae in the most general case, which have
applications in multivariate statistics. The reader interested in
proofs is referred to \cite{DiNardo3}.

The starting point is the expression of moments of $n.(\chi \alpha)$ in
terms of $n.\alpha$ and viceversa:
\begin{eqnarray}
[n.(\chi \alpha)]^i & \simeq &   \sum_{\lambda \vdash i}
d_{\lambda} (\chi.\chi)_{\lambda} (n.\alpha)^{r_1}
(n.\alpha^2)^{r_2} \cdots,
\label{(augpow)} \\
(n.\alpha)^i & \simeq & \sum_{\lambda \vdash i} d_{\lambda}
[n.(\chi\alpha)]^{r_1}[n.(\chi\alpha^2)]^{r_2}\cdots.
\label{(momdot2)}
\end{eqnarray}
Such equivalences involve integer partitions and are very easy to
implement since there is at least one procedure devoted to integer partitions
in any symbolic package. Note that equivalences (\ref{(augpow)}) and
(\ref{(momdot2)})  may be rewritten replacing $\alpha$ with any power
$\alpha^k.$ For instance, in (\ref{(momdot2)}) we have
$$(n.\alpha^k)^i  \simeq   \sum_{\lambda \vdash i} d_{\lambda}
[n.(\chi\alpha^k)]^{r_1}[n.(\chi\alpha^{2k})]^{r_2}\cdots.
$$
The next step is to express more general products
$[n.(\chi\alpha)]^{r_1}[n.(\chi\alpha^2)]^{r_2}\cdots$ (that is
augmented symmetric polynomials) in terms of power sums. With this
aim, equivalences (\ref{(augpow)}) and (\ref{(momdot2)}) must be
rewritten by using set partitions instead of integer partitions.
We say in advance that the final step will consist in replacing
the set with the more general structure of multiset.
\par
Let $C$ be a subset of $R[A]$ with $n$ elements. Recall that a
partition $\pi$ of $C$ is a collection $\pi=\{B_1, B_2, \ldots,
B_k\}$ with $k \leq n$ disjoint and not-empty subsets of $C$ whose
union is $C.$ We denote by $\Pi_n$ the set of all partitions of
$C.$ Let $\{\alpha_1,\alpha_2,\ldots,\alpha_n\}$ be a set of $n$
uncorrelated umbrae similar to an umbra $\alpha.$ The symbol
$\alpha^{.\pi}$ denotes the umbra
\begin{equation}
\alpha^{.\pi} \equiv \alpha_{i_1}^{|B_1|} \alpha_{i_2}^{|B_2|}
\cdots \alpha_{i_k}^{|B_k|}, \label{(1.1)}
\end{equation}
where $\pi=\{B_1,B_2, \ldots, B_k\}$ is a partition of
$\{\alpha_1,\alpha_2,\ldots,\alpha_n\}$ and $i_1, i_2, \ldots,
i_k$ are distinct integers chosen in $\{1,2,\ldots,n\}.$ Note that
$\alpha^{.\pi} \equiv \alpha_{\lambda},$ when $\lambda$ is the
partition of the integer $n$ determined by $\pi.$ Indeed, a set
partition is said to be of type $\lambda=(1^{r_1},2^{r_2},\ldots)$
if there are $r_1$ blocks of cardinality $1,$ $r_2$ blocks of
cardinality $2$ and so on. The number of set partitions of type
$\lambda$ is $d_{\lambda},$ as given in (\ref{(not)}). By using
set partitions, equivalences (\ref{(augpow)}) and
(\ref{(momdot2)}) may be rewritten as
\begin{eqnarray}
(n.\alpha)^i & \simeq &  \sum_{\pi \in \Pi_i} [n.(\chi
\alpha)]^{r_1} [n.(\chi \alpha^2)]^{r_2} \cdots, \label{(bis1)} \\
{[n.(\chi \alpha)]^i} & \simeq &  \sum_{\pi \in \Pi_i}
(\chi.\chi)^{.\pi} (n.\alpha)^{r_1} (n.\alpha^2)^{r_2} \cdots,
\label{(ter1)}
\end{eqnarray}
where $ E[(\chi.\chi)^{.\pi}]= E \left[(\chi.\chi)^{|B_1|}
(\chi^{\prime}.\chi^{\prime})^{|B_2|} \cdots (\chi^{\prime
\prime}.\chi^{\prime \prime})^{|B_k|}\right] = \prod_{i=1}^k
x_{|B_i|}$ from (\ref{(1.1)}). From a computational point of view,
equivalences (\ref{(bis1)}) and (\ref{(ter1)}) are  less efficient
than equivalences (\ref{(augpow)}) and (\ref{(momdot2)}). The
computational cost is $O(B_n),$ where $B_n$ is the $n$-th Bell
number whose growth is greater than $e^n.$ Anyway, equivalences
(\ref{(bis1)}) and (\ref{(ter1)}) smooth the way to generalize
such computations to the multivariate case, by using the notion of
multiset.
\par
A  multiset $M$ is a pair $(\bar{M}, f),$ where $\bar{M} \subset
R[A]$ is a set, called the support of the multiset, and $f$ is a
function from $\bar{M}$ to the non-negative integers. For each
$\mu \in \bar{M},$ $f(\mu)$ is called the multiplicity of $\mu.$
If the support of $M$ is a finite set, say $\bar{M}=\{\mu_1,
\mu_2, \ldots, \mu_k\},$ we write
$$M = \{\mu_1^{(f(\mu_1))},\mu_2^{(f(\mu_2))}, \ldots, \mu_k^{(f
(\mu_k))}\} \quad \hbox{or} \quad M = \{\underbrace{\mu_1, \ldots,
\mu_1}_{f(\mu_1)}, \ldots, \underbrace{\mu_k, \ldots,
\mu_k}_{f(\mu_k)}\}.$$ The length of the multiset $M$ is the sum
of multiplicities of all elements of $\bar{M},$ that is
$$|M| = \sum_{\mu \in \bar{M}} f(\mu).$$
From now on, we denote a multiset $(\bar{M}, f)$ simply by $M.$
For instance the multiset
$$M = \{\underbrace{\alpha, \ldots, \alpha}_{i}\} = \{\alpha^{(i)}\}$$
has length $i,$ support $\bar{M} = \left\{ \alpha \right\}$ and
$f(\alpha)=i.$ In the following, we set
\begin{equation}
\mu_{M} =  \prod_{\mu \in \bar{M}} \mu^{f(\mu)} \quad \hbox{and} \quad
(n.\mu)_{M} =  \prod_{\mu \in \bar{M}} (n.\mu)^{\,g(\mu)}. \label{(momb4)}
\end{equation}
For instance, if $M = \{\alpha^{(i)}\}$ then $(n.\alpha)_M \simeq
(n.\alpha)^i$ and $[n.(\chi \alpha)]_M \simeq [n.(\chi
\alpha)]^i.$ Note that this notation can be easily extended to
umbral polynomials.

If $\lambda=(1^{r_1},2^{r_2},\ldots)$ is an integer partition, set
\begin{equation}
P_{\lambda} = \{\underbrace{\alpha, \ldots, \alpha}_{r_1}, \underbrace{\alpha^2, \ldots, \alpha^2}_{r_2}, \ldots \}.
\label{(set)}
\end{equation}
By using the notation (\ref{(momb4)}), we have
$(n.\alpha)_{P_{\lambda}} \simeq  (n.\alpha)^{r_1}
(n.\alpha^2)^{r_2} \cdots$ and ${[n.(\chi \alpha)]}_{P_{\lambda}}
\simeq [n.(\chi\alpha)]^{r_1} [n.(\chi\alpha^2)]^{r_2} \cdots,$ so
equivalences (\ref{(bis1)}) and (\ref{(ter1)}) may be more
compressed
\begin{equation}
(n.\alpha)_M \simeq  \sum_{\pi \in \Pi_{i}} [n.(\chi\alpha)]_{P_{\lambda}} \quad \hbox{and} \quad
[n.(\chi \alpha)]_M \simeq  \sum_{\pi \in \Pi_{i}} (\chi.\chi)^{.\pi} (n.\alpha)_{P_{\lambda}},
\label{(ter2)}
\end{equation}
where $\lambda$ is the type of the set partition $\pi.$

In equivalences (\ref{(ter2)}), we have $M=\{ \alpha^{(i)} \}.$
The last step consists in generalizing such equivalences to any
multiset $M.$ To this aim, we recall the notion of multiset
subdivision. Such a notion is quite natural and it is equivalent to
splitting the multiset into disjoint blocks (submultisets) whose
union gives the whole multiset.

A subdivision of a multiset $M$ is a multiset $S=(\bar{S},g)$ of
$k \leq |M|$ non-empty submultisets $M_i=(\bar{M}_i, f_i)$ of $M$
such that
\begin{description}
\item[{\it i)}] $\cup_{i=1}^k \bar{M}_i = \bar{M};$ \item[{\it
ii)}] $\sum_{i=1}^k f_i(\mu) = f(\mu)$ for any $\mu \in \bar{M}.$
\end{description}
Recall that a multiset $M_i=(\bar{M_i},f_i)$ is a submultiset of $M=(\bar
{M}, f)$ if $\bar{M_i} \subseteq \bar{M}$ and $f_i(\mu) \leq f(\mu), \,
\forall \mu \in \bar{M_i}.$

If $M=\{\alpha^{(i)}\},$ then subdivisions are of type
$$S=\{\underbrace{\{\alpha\},\ldots,\{\alpha\}}_{r_1}, \underbrace{\{\alpha^{(2)}\},\ldots,\{\alpha^{(2)}\}}_{r_2}, \ldots\}$$
with $r_1 + 2 \,r_2 + \cdots = i,$ and we will say that the subdivision $S$ is of type $\lambda = (1^{r_1}, 2^{r_2}, \ldots)
\vdash i.$ The support of $S$ is $\bar{S}=\{\{\alpha\}, \{\alpha^{(2)}\}, \ldots\}.$

If
\begin{equation}
S=\{\underbrace{M_1 \ldots, M_1}_{g(M_1)}, \underbrace{M_2,\ldots,M_2}_{g(M_2)}, \ldots,
\underbrace{M_k,\ldots,M_k}_{g(M_k)}\},
\label{(typesub)}
\end{equation}
we set
\begin{equation}
\mu_{S} =  \prod_{M_i \in \bar{S}} \mu_{M_i}^{g(M_i)} \quad
\hbox{and} \quad (n.\mu)_{S} =  \prod_{M_i \in \bar{S}}
(n.\mu_{M_i})^{\,g(M_i)}, \label{(momb6)}
\end{equation}
extending the notation (\ref{(momb4)}).

When integer partitions are replaced by multiset subdivisions, the
fundamental expectation result (\ref{(ustat)}) becomes
\begin{equation}
[n.(\chi\mu)]_{S}  \simeq  (n)_{|S|} \mu^{.S},
\label{(ustat1)}
\end{equation}
with $S$ given in (\ref{(typesub)}) and
\begin{equation}
\mu^{.S} \equiv (\mu_{M_1})^{.g(M_1)} \cdots
(\mu^{\prime}_{M_k})^{.g(M_k)}, \label{(momb6bis)}
\end{equation}
where $\mu_{M_t}$ are uncorrelated umbral monomials.

By using the notation (\ref{(momb6)}) and recalling (\ref{(set)}),
we have $(n.\alpha)_{P_{\lambda}} \equiv (n.\alpha)_S$ and
$[n.(\chi\alpha)]_{P_{\lambda}} \equiv [n.(\chi\alpha)]_S,$ with
$\lambda=(1^{r_1},2^{r_2},\ldots) \vdash i$ and $S$ the
subdivision of type $\lambda.$ Then equivalences (\ref{(ter2)})
may be written as follows
\begin{equation}
(n.\alpha)_M \simeq  \sum_{\pi \in \Pi_{i}} [n.(\chi\alpha)]_{S}
\quad \hbox{and} \quad [n.(\chi \alpha)]_M \simeq  \sum_{\pi \in
\Pi_{i}} (\chi.\chi)^{.\pi} (n.\alpha)_{S}. \label{(ter33)}
\end{equation}

One more remark allows us to remove integer partitions from
(\ref{(ter33)}) which is necessary when the multiset
$M=\{\alpha^{(i)}\}$ is replaced by an arbitrary multiset. Let us
observe that a subdivision of the multiset $M$ may be constructed
in the following way: suppose the elements of $M $ to be all
distinct, build a set partition and then replace each element in
any block by the original one. In this way, any subdivision
corresponds to a set partition $\pi$ and we will write $S_{\pi}.$
Note that it is $|S_{\pi}|=|\pi|$ and it could be $S_{\pi_1} =
S_{\pi_2}$ for $\pi_1 \ne \pi_2,$ as the following example shows.
\begin{example}{\rm If  $M=\{ \alpha, \alpha, \gamma, \delta, \delta\},$
label each element of $M$ in order to have the set $C=\{ \alpha_1,
\alpha_2, \gamma_1, \delta_1, \delta_2\}.$ The subdivision
$S_1=\{\{\alpha,\gamma\},\{\alpha\}, \{\delta,\delta\}\}$
corresponds to the partition $\pi_1=
\{\{\alpha_1,\gamma_1\},\{\alpha_2\}, \{\delta_1,\delta_2\}\}$ of
$C.$ It is $|S_{1}|=|\pi_1|.$ Note that the subdivision $S_1$ also
corresponds to the partition $\pi_2=
\{\{\alpha_2,\gamma_1\},\{\alpha_1\}, \{\delta_1,\delta_2\}\}.$  $\Box$}
\end{example}

Finally, equivalences (\ref{(ter2)}) may be rewritten as follows
\begin{equation}
(n.\mu)_M \simeq  \sum_{\pi \in \Pi_{i}} [n.(\chi\mu)]_{S_{\pi}} \quad \hbox{and} \quad
[n.(\chi \mu)]_M \simeq  \sum_{\pi \in \Pi_{i}} (\chi.\chi)^{.\pi} (n.\mu)_{{S_{\pi}}},
\label{(ter3)}
\end{equation}
where $S_{\pi}$ is the subdivision of $M$ corresponding to the
partition $\pi$ of a set $C$ such that $|C|=|M|=i.$ The symbolic
expression of such equivalences does not change if the multiset $M
= \{\alpha^{(i)}\}$ is replaced by an arbitrary multiset.

The following example shows the effectiveness of umbral notation
in managing moments of sampling distributions.
\begin{example} \label{ex2}
{\rm If $M=\{ \mu_1, \mu_1, \mu_2\}$ then $(n.\mu)_M = (n.\mu_1)^2
(n.\mu_2).$ In statistical terminology, moments of $(n.\mu)_M$
correspond to moments of the product of sums $\left( \sum_{i=1}^n
X_{i} \right)^2 \left( \sum_{i=1}^n Y_{i} \right),$ where
$(X_1,Y_1), \ldots, (X_n,Y_n)$ are separately independent and
identically distributed random variables. In order to apply the
first part of (\ref{(ter3)}), we need to compute all subdivisions
of $M.$ These are given in Table 1.
\begin{table}[h]
\begin{center}
\begin{tabular}{|c||c||c|} \hline & & \\[-10pt]
$ S_{\pi} $ & $\sharp S_{\pi}$ & $[n.(\chi \mu)]_{S_{\pi}}$ \\[3pt] \hline & &  \\[-10pt]
$\{\{\mu_1, \mu_1, \mu_2\}\}$ & $1$ & $ n.(\chi \mu_1^2 \mu_2)$ \\[3pt] \hline & &  \\[-10pt]
$\{\{\mu_1\},\{\mu_1,\mu_2\}\}$ & $2$ & $[n.(\chi \mu_1)] [n.(\chi \mu_1 \mu_2)]$ \\[3pt] \hline & &  \\[-10pt]
$\{\{\mu_2\},\{\mu_1,\mu_1\}\}$ & $1$ & $[n.(\chi \mu_2)] [n.(\chi \mu_1^2)]$ \\[3pt] \hline & &  \\[-10pt]
$\{\{\mu_1\},\{\mu_1\},\{\mu_2\}\}$ & $1$ &   $[n.(\chi \mu_1)]^2
[n.(\chi \mu_2)]$      \\[3pt] \hline
\end{tabular}
\\ \medskip{Table 1: Subdivisions of  $M=\{ \mu_1, \mu_1, \mu_2\}.$}
\end{center}
\end{table}

The last column in Table 1 has been constructed by the following
considerations. Suppose to consider the second row:
$S_1=\{\{\mu_1\},\{\mu_1,\mu_2\}\}$ is a subdivision of $M = \{
\mu_1, \mu_1, \mu_2\}.$ The support of $S_1$ consists of two
multisets, $M_{1}=\{\mu_1\}$ and $M_{2}=\{\mu_1, \mu_2\},$ each of
one with multiplicity $1$ so that $[n.(\chi\mu)]_{S_{1}} = n.(\chi
\mu_{M_1}) n.(\chi\mu_{M_2}).$ Since $n.(\chi \mu_{M_1}) = n.(\chi
\mu_1)$ and $n.(\chi\mu_{M_2}) =  n.(\chi \mu_1 \mu_2),$ we have
$[n.(\chi\mu)]_{S_{1}}= n.(\chi \mu_1) \,\, n.(\chi \mu_1 \mu_2).$
Repeating the same arguments for all subdivisions of $M,$ we get
the results of Table 1. By using Table 1 and the first part of
(\ref{(ter3)}), we have
\begin{eqnarray*}
(n.\mu_1)^2 (n.\mu_2) & \simeq & n.(\chi \mu_1^2 \mu_2) + 2 [n.(\chi \mu_1)] [n.(\chi \mu_1 \mu_2)] \\
& + & [n.(\chi \mu_2)] [n.(\chi \mu_1^2)] + [n.(\chi \mu_1)]^2 [n.(\chi \mu_2)],
\end{eqnarray*}
that is
\begin{eqnarray}
\left( \sum_{i=1}^n X_{i} \right)^2 \left( \sum_{i=1}^n Y_{i} \right)& = &  \sum_{i=1}^n
X^2_{i} Y_{i}  + 2 \sum_{1 \leq i \ne j \leq n} X_{i} X_{j} Y_{j} \nonumber \\
& + &  \sum_{1 \leq i \ne j \leq n} X^2_{i} Y_{j}  +  \sum_{1 \leq i \ne j \ne k \leq n} X_{i} X_{j} Y_{k}.
\label{(exx1)}
\end{eqnarray}
In order to evaluate the mean of the sums, on the right hand side of (\ref{(exx1)}), in terms of population moments,
we have to use (\ref{(ustat1)}) and finally we have
\begin{eqnarray*}
E\left[ \left( \sum_{i=1}^n X_{i} \right)^2 \left( \sum_{i=1}^n Y_{i} \right) \right] & = & n E[
X^2 Y]  + 2 (n)_2 E[ X]E[ X Y] \\
& + &  (n)_2 E[X^2]E[Y]  +  (n)_3 E[X]^2 E[Y].
\end{eqnarray*} $\Box$}
\end{example}
In the previous example, we have shown the usefulness of the first
part of (\ref{(ter3)}) in evaluating the mean of product of power
sums. In the next example, we show the usefulness of the second
part of (\ref{(ter3)}) in order to express any $U$-statistic in
terms of power sums. Indeed, first we make use of equivalence
(\ref{(ustat1)}), in order to translate products of moments - also
multivariate - in terms of augmented symmetric polynomials. Then
we apply the change of bases given by the second part of
(\ref{(ter3)}).

The following example shows how to construct multivariate $k$-statistics.

\begin{example}{\it Multivariate $k$-statistics.}
{\rm In umbral terms, a  multivariate cumulant is the element of
$R$ corresponding to $E[(\chi.\mu)_M]= \kappa_{t_1 \ldots t_r},$
where $M = \{\mu_1^{(t_1)}, \mu_2^{(t_2)}, \ldots,
\mu_r^{(t_r)}\}$ is a multiset and $(\chi.\mu)_M$ is the symbol
denoting the product $(\chi.\mu_1)^{t_1} (\chi.\mu_2)^{t_2} \cdots
(\chi.\mu_r)^{t_r}.$ Suppose $|M|=i.$ The extension of
(\ref{(cum)}) to the multivariate case is
$$(\chi.\mu)_{M} \simeq  \sum_{\pi \in \Pi_i} (\chi.\chi)^{|\pi|} \,
\mu^{.{S_\pi}},$$
where $S_{\pi}$ is the subdivision of the multiset $M$ corresponding to the partition $\pi \in \Pi_{i}.$
By equivalence (\ref{(ustat1)}), we write
$$(\chi.\mu)_{M} \simeq  \sum_{\pi \in \Pi_{i}} (\chi.\chi)^{|\pi|} \,
\frac{1}{(n)_{|\pi|}} [n.(\chi\mu)]_{S_{\pi}},$$ and by the second
part of (\ref{(ter3)}) we have the umbral version of multivariate
$k$-statistics in terms of power sums, that is
\begin{equation}
(\chi.\mu)_{M} \simeq  \sum_{\pi \in \Pi_{k}}
\frac{(\chi.\chi)^{|\pi|} }{(n)_{|\pi|}} \, \sum_{\tau \in
\Pi_{|\pi|}} (\chi.\chi)^{.\tau} (n.\mu)_{S_{\tau}},
\label{(dotprod12)}
\end{equation}
where $S_{\tau}$ is the subdivision of $M$ corresponding to the
partition $\tau$ of a set having the same cardinality of $\pi.$  $\Box$}
\end{example}
\section{Products of augmented polynomials}
This section is devoted to a different application of equivalences
(\ref{(ter3)}), necessary to evaluate the mean of product of
augmented polynomials in separately independent and identically
distributed random variables.

We borrow the following example from the paper of Vrbik \cite{Vrbik}.

Suppose, for instance, to need the mean of
\begin{equation}
\left( \sum_{i\ne j}^n X_{i}^2 \, X_{j} \right) \, \left(
\sum_{i=1}^n X_{i}^2 \, Y_i \right)^2 = S_{\{\{2,0\}\,,\{1,0\}\}}
\,S_{\{\{2,1\}\}} \,S_{\{\{2,1\}\}}, \label{(sample)}
\end{equation}
where $(X_1,Y_1), \ldots, (X_n,Y_n)$ are separately independent
and identically distributed random variables and
\begin{equation}
S_{\{\{k_1,\, l_1\}\,,\{k_2,\, l_2\},\ldots\}} = \sum_{i_1 \ne i_2
\ne \cdots} X_{i_1}^{k_1} \, Y_{i_1}^{l_1} \, X_{i_2}^{k_2} \,
Y_{i_2}^{l_2} \cdots, \label{(notation)}
\end{equation}
by means of the notation introduced by Vrbik in \cite{Vrbik}.

If we expand the product (\ref{(sample)}) as a linear combination of augmented
symmetric polynomials (\ref{(notation)}), then we are able  to apply the
fundamental expectation result (\ref{(ustat1)}) and to
evaluate the mean of (\ref{(sample)}). The umbral tools, we have
introduced up to now, are sufficient to do such a work.
Therefore, since
\begin{eqnarray*}
\left( \sum_{i\ne j}^n X_{i}^2 X_{j} \right) \quad & \hbox{corresponds to} & \quad
[n.(\chi \mu_1^2) n.(\chi \mu_1)], \quad \hbox{and} \\
\left( \sum_{i=1}^n X_{i}^2 Y_i \right) \quad & \hbox{corresponds to} &  \quad
[n.(\chi \mu_1^2 \mu_2)],
\end{eqnarray*}
the product (\ref{(sample)}) is umbrally represented by
\begin{equation}
n.(\chi_1 \mu_1^2) \, n.(\chi_1 \mu_1) \, n.(\chi_2 \mu_1^2 \mu_2) \,
n.(\chi_3 \mu_1^2 \mu_2),
\label{(sample2)}
\end{equation}
where $\{\chi_i\},i=1,2,3$ are uncorrelated singleton umbrae.
Indeed, a product of uncorrelated singleton umbrae does not \lq\lq
delete\rq\rq \, the same indexed umbrae. Moreover, the sum
(\ref{(notation)}) is umbrally represented by
\begin{equation}
n.(\chi \mu_1^{k_1} \mu_2^{l_1}) \, n.(\chi \mu_1^{k_2} \mu_2^{l_2}) \cdots.
\label{(sample31)}
\end{equation}

\begin{center}
\begin{tabular}{|c||c|} \hline & \\[-10pt]
$ S_{\pi} $ & $S_{\{\{k_1,l_1\},\{k_2,l_2\},\ldots\}}$ \\[3pt] \hline & \\[-10pt]
$\{ \{\chi_1 \mu_1^2\}, \{\chi_1 \mu_1\},\{ \chi_2 \mu_1^2
\mu_2\}, \{\chi_3 \mu_1^2 \mu_2\} \}$ &
$S_{\{\{2,0\},\{1,0\},\{2,1\},\{2,1\}\}}$ \\[3pt] \hline & \\[-10pt] $\{ \{\chi_1
\mu_1^2, \chi_1 \mu_1\},\{\chi_2 \mu_1^2 \mu_2\},\{\chi_3 \mu_1^2
\mu_2\} \}$ & $0$ \\[3pt] \hline & \\[-10pt] $\{ \{\chi_1 \mu_1^2, \chi_2 \mu_1^2
\mu_2\},\{\chi_1 \mu_1\}, \{\chi_3 \mu_1^2 \mu_2\} \}$ &
$S_{\{\{4,1\},\{1,0\},\{2,1\}\}}$ \\[3pt] \hline & \\[-10pt] $\{ \{\chi_1 \mu_1^2,
\chi_3 \mu_1^2 \mu_2\},\{\chi_2 \mu_1^2 \mu_2\}, \{\chi_1 \mu_1\}
\}$ & $S_{\{\{4,1\},\{1,0\},\{2,1\}\}}$ \\[3pt] \hline & \\[-10pt] $\{ \{\chi_1
\mu_1, \chi_2 \mu_1^2 \mu_2 \},\{\chi_1 \mu_1^2\}, \{\chi_3
\mu_1^2 \mu_2\} \}$ & $S_{\{\{3,1\},\{2,0\},\{2,1\}\}}$  \\[3pt] \hline & \\[-10pt]
$\{ \{\chi_1 \mu_1, \chi_3 \mu_1^2 \mu_2\}, \{\chi_1 \mu_1^2\},
\{\chi_2 \mu_1^2 \mu_2\} \}$ & $S_{\{\{3,1\},\{2,0\},\{2,1\}\}}$
\\[3pt] \hline & \\[-10pt] $\{ \{\chi_2 \mu_1^2 \mu_2,\chi_3 \mu_1^2
\mu_2\},\{\chi_1 \mu_1\}, \{\chi_1 \mu_1^2\} \}$ &
$S_{\{\{4,2\},\{1,0\},\{2,0\}\}}$  \\[3pt] \hline & \\[-10pt] $\{ \{\chi_1 \mu_1^2,
\chi_1 \mu_1\},\{\chi_2 \mu_1^2 \mu_2, \chi_3 \mu_1^2 \mu_2\} \}$
& $0$  \\[3pt] \hline & \\[-10pt] $\{ \{\chi_1 \mu_1^2, \chi_2 \mu_1^2
\mu_2\},\{\chi_1 \mu_1, \chi_3 \mu_1^2 \mu_2\} \}$ &
$S_{\{\{4,1\},\{3,1\}\}}$ \\[3pt] \hline & \\[-10pt] $\{ \{\chi_1 \mu_1^2, \chi_3
\mu_1^2 \mu_2\},\{\chi_2 \mu_1^2 \mu_2, \chi_1 \mu_1\} \}$ &
$S_{\{\{4,1\},\{3,1\}\}}$  \\[3pt] \hline & \\[-10pt] $\{ \{\chi_1 \mu_1, \chi_2
\mu_1^2 \mu_2, \chi_1 \mu_1^2\}, \{\chi_3 \mu_1^2 \mu_2\} \}$ &
$0$ \\ \hline $\{ \{\chi_1 \mu_1, \chi_3 \mu_1^2 \mu_2, \chi_1
\mu_1^2\}, \{\chi_2 \mu_1^2 \mu_2\} \}$ & $0$ \\[3pt] \hline & \\[-10pt] $\{
\{\chi_2 \mu_1^2 \mu_2,\chi_3 \mu_1^2 \mu_2, \chi_1 \mu_1\},
\{\chi_1 \mu_1^2\} \}$ & $S_{\{\{5,2\},\{2,0\}\}}$ \\[3pt] \hline & \\[-10pt] $\{
\{\chi_2 \mu_1^2 \mu_2,\chi_3 \mu_1^2 \mu_2, \chi_1 \mu_1^2\},
\{\chi_1 \mu_1\} \}$ & $S_{\{\{6,2\},\{1,0\}\}}$ \\[3pt] \hline & \\[-10pt] $\{
\{\chi_1 \mu_1^2, \chi_1 \mu_1, \chi_2 \mu_1^2 \mu_2, \chi_3
\mu_1^2 \mu_2\} \}$ & $0$   \\[3pt] \hline
\end{tabular}
\\ \medskip{Table 2: Subdivisions of  $M=\{ \chi_1 \mu_1^2, \chi_1 \mu_1, \chi_2 \mu_1^2 \mu_2, \chi_3 \mu_1^2 \mu_2\}.$}
\end{center}

Let $M=\{ \chi_1 \mu_1^2, \chi_1 \mu_1, \chi_2 \mu_1^2 \mu_2,
\chi_3 \mu_1^2 \mu_2\}.$ The length of $M$ is $4,$ and since the
monomials are all different, $M$ is a set. From the second part of
(\ref{(momb4)}), we have
\begin{equation}
(n.\mu)_M \simeq [n.(\chi_1 \mu_1^2) n.(\chi_1 \mu_1)] [n.(\chi_2 \mu_1^2 \mu_2)]
[n.(\chi_3 \mu_1^2 \mu_2)].
\label{(sample3)}
\end{equation}
Via the first part of equivalence (\ref{(ter3)}), we have
\begin{equation}
[n.(\chi_1 \mu_1^2) n.(\chi_1 \mu_1)] [n.(\chi_2 \mu_1^2 \mu_2)]
[n.(\chi_3 \mu_1^2 \mu_2)] \simeq \sum_{\pi \in \Pi_{4}} [n.(\chi \mu)]_{\pi},
\label{eee}
\end{equation}
where $S_{\pi} = \pi,$ since $M$ is a set, $\Pi_4$ is the set of
all partitions of $M$ and $\mu$ is an element of $M.$ We must pay
attention to the auxiliary umbra $[n.(\chi \mu)]_{\pi},$ whose
structure looks like (\ref{(sample31)}). For example, let us
consider the partition $\pi_1=\{ \{ \chi_1 \mu_1^2, \chi_2 \mu_1^2
\mu_2\}, \{\chi_1 \mu_1, \chi_3 \mu_1^2 \mu_2\}\}.$ Observe that
$[n.(\chi \mu)]_{\pi_1} = n.(\chi_1 \chi_2 \mu_1^4 \mu_2)$ $n.
(\chi_1 \chi_3 \mu_1^3 \mu_2) \simeq n.(\chi \mu_1^4 \mu_2) \, n.
(\chi \mu_1^3 \mu_2),$ corresponding to $S_{\{\{4,1\},\{3,1\}\}}$
in (\ref{(notation)}). If we consider $\pi_2=\left\{ \{ \chi_1
\mu_1^2, \chi_1 \mu_1\}, \right.$ $\left\{\chi_2 \mu_1^2 \mu_2,
\chi_3 \mu_1^2 \mu_2\}\right\},$ we have $[n.(\chi \mu)]_{\pi_2} =
n.(\chi \, \chi_1^2 \, \mu_1^3) \, n. (\chi \, \chi_2 \chi_3 \,
\mu_1^4 \mu_2^2).$  We have $n. (\chi \, \chi_2 \chi_3 \, \mu_1^4
\mu_2^2)  \simeq n. (\chi \mu_1^4 \mu_2^2)$ with the deletion of
indexed singleton umbrae, but $n.(\chi \, \chi_1^2 \, \mu_1^3)
\simeq 0,$ since $\chi_1^2 \simeq 0.$ So we have $[n.(\chi
\mu)]_{\pi_2} \simeq 0.$

In conclusion, when in one - or more - blocks of the subdivision
$S_{\pi}$, there are at least two umbral monomials involving
correlated singleton umbrae, the auxiliary umbra $[n.(\chi
\mu)]_{S_{\pi}}$ has the evaluation equal to zero. If within every
block of the subdivision $S_{\pi}$ there are only uncorrelated
singleton umbrae, then $[n.(\chi \mu)]_{S_{\pi}}$ gives rise
expressions like (\ref{(sample31)}), umbrally representing
(\ref{(notation)}).

We do all computation in Table 2. We give the corresponding sum of
independent random variables (\ref{(notation)}), instead of
$[n.(\chi \mu)]_{S_{\pi}}.$ Subsection 6.1 is devoted to the
algorithm {\tt makeTab}, which allows us to construct multiset
subdivisions similar to those in the first column of Table 1 and
Table 2.
\section{Computational results}
\subsection{Find subdivisions of a multiset: the procedure {\tt makeTab}}
Equivalences (\ref{(ter3)}) have been implemented in {\tt MAPLE}.
These equivalences share the procedure {\tt makeTab} necessary to
construct multiset subdivisions.

When the multiset is of type $\{\alpha^{(k)}\}$, an efficient way
is to resort the partitions of the integer $k,$ as equivalences
(\ref{(ter2)}) show. In general, as already stressed in Section 4,
we may construct multiset subdivisions by using suitable set
partitions, but this approach has a computational cost
proportional to the $n$-th Bell number $B_n,$ so it is not
efficient. Indeed, examples have shown how subdivisions may occur
more than one time in the same formula (see Table 1), so that it
is necessary to build a procedure generating only different
subdivisions together with their multiplicity, that is the number
of corresponding set partitions. To accomplish this task, the
algorithm {\tt makeTab} takes into account the connection between
multisets and integer partitions, reducing the overall
computational complexity.

In the following, we illustrate the main steps of {\tt makeTab} by
an example. Suppose to need  subdivisions of the multiset
$$M=\{\underbrace{\alpha, \alpha, \alpha}_3, \underbrace{\gamma, \gamma}_2\}.$$
We compute all different subdivisions of $\{\alpha^{(3)}\}$ by
using all partitions $\lambda$ of the interger $3,$ that is
\begin{equation}
\{\{\alpha\},\{\alpha\},\{\alpha\}\}, \quad \{\{\alpha\}, \{\alpha^{(2)}\}\}, \quad
\{\{\alpha^{(3)}\}\}.
\label{(sub1)}
\end{equation}
The same we do for $\{\gamma^{(2)}\},$ that is
\begin{equation}
\{\{\gamma\},\{\gamma\}\}, \quad \{\{\gamma^{(2)}\}\}.
\label{(subgamma1)}
\end{equation}

Now, we {\it insert} every element of (\ref{(subgamma1)}) in every
element of (\ref{(sub1)}) one at a time and recursively, as the
following example shows. Suppose to do the insertion of
$\{\{\gamma\},\{\gamma\}\}$ in
$\{\{\alpha\},\{\alpha\},\{\alpha\}\}.$ We first insert
$\{\gamma\}$ in every block of
$\{\{\alpha\},\{\alpha\},\{\alpha\}\},$ that is
$$\{\{\alpha\},\{\alpha\},\{\alpha\}\} \leftarrow \gamma.$$
Then, we insert a second time $\{\gamma\}$ in the output
subdivision, that is
$$\Bigl( \{\{\alpha\},\{\alpha\},\{\alpha\}\} \leftarrow \{\gamma\} \Bigr)  \leftarrow \{\gamma\}.$$
The insertion $\leftarrow$ is a kind of iterated
inclusion-exclusion rule \cite{Andrews1}, but with some more
constraints:
\begin{description}
\item[{\it i)}] the insertion of a submultiset of
$\{\gamma^{(2)}\}$ in a submultiset of $\{\alpha^{(3)}\}$ must be
done only if it does not generate a new submultiset equal to a
previous one or it has not yet inserted; \item[{\it ii)}] at the
end, every submultiset of $\{\gamma^{(2)}\}$ is simply appended to
every subdivision of $\{\alpha^{(3)}\}.$
\end{description}
Table 3 gives the results of the double insertion of
$\{\{\gamma\},\{\gamma\}\}$ in every submultiset of
$\{\alpha^{(3)}\},$ according to rules {\it i)} and {\it ii)}.

\begin{center}
\begin{tabular}{|l||l||l|} \hline & & \\[-10pt]
Subdivision &  Output first $\leftarrow$ & Output second $\leftarrow$  \\[3pt] \hline & & \\[-10pt]
$\{\{\alpha\},\{\alpha\},\{\alpha\}\}$ & $\{\{\alpha, \gamma \},\{\alpha\},\{\alpha\}\}$   & $\{\{\alpha, \gamma \},\{\alpha, \gamma\},\{\alpha\}\}$
\\[3pt] \hline & & \\[-10pt]
                                       & $\{\{\alpha\},\{\alpha\},\{\alpha\},\{\gamma\}\}$ & $\{\{\alpha, \gamma \},\{\alpha\},\{\alpha\},\{\gamma\}\}$
\\[3pt]
                                       &                                                   & $\{\{\alpha\},\{\alpha\},\{\alpha\},\{\gamma\},\{\gamma\}\}$
\\[3pt] \hline & & \\[-10pt]
$\{ \{\alpha\},\{\alpha^{(2)}\}\}$     & $\{ \{\alpha, \gamma\},\{\alpha^{(2)} \}\}$       & $\{ \{\alpha, \gamma\},\{\alpha^{(2)}, \gamma \}\}$\\[3pt]
                                       & $\{ \{\alpha \},\{\alpha^{(2)}, \gamma\}\}$       & $\{ \{\alpha, \gamma\},\{\alpha^{(2)} \}, \{\gamma\}\}$ \\[3pt]
                                       & $\{ \{\alpha \},\{\alpha^{(2)}\}, \{\gamma\}\}$   & $\{ \{\alpha \},\{\alpha^{(2)}, \gamma\}, \{\gamma\}\}$ \\[3pt]
                                       &                                                   & $\{ \{\alpha \},\{\alpha^{(2)}\}, \{\gamma\}\}$ \\[3pt] \hline & & \\[-10pt]
$\{ \{\alpha^{(3)}\}\}$                & $\{ \{\alpha^{(3)}, \gamma \}\}$                  & $\{ \{\alpha^{(3)} \gamma, \gamma \}\}$ \\[3pt]
                                       & $\{ \{\alpha^{(3)}\}, \{\gamma \}\}$              & $\{ \{\alpha^{(3)}\}, \{\gamma \}, \{\gamma\}\}$ \\[3pt]
\hline
\end{tabular}
\\ \smallskip{Table 3: Insertion of $\{\{\gamma\},\{\gamma\}\}$ in every submultiset
of $\{\alpha^{(3)}\}$}
\end{center}
Table 4 gives the results of the insertion of
$\{\{\gamma^{(2)}\}\}$ in every submultiset of $\{\alpha^{(3)}\},$
according to rules {\it i)} and {\it ii)}.
\begin{center}
\begin{tabular}{|l||l|} \hline  & \\[-10pt]
Subdivision &  Output first $\leftarrow$ \\[3pt] \hline & \\[-10pt]
$\{\{\alpha\},\{\alpha\},\{\alpha\}\} \leftarrow \{\gamma^{(2)}\}$ &
$\{\{\alpha,\gamma^{(2)}\},\{\alpha\},\{\alpha\}\}$  $\{\{\alpha\},\{\alpha\},\{\alpha\},\{\gamma^{(2)}\}\} $  \\[3pt] \hline &  \\[-10pt]
$\{ \{\alpha\},\{\alpha^{(2)}\}\} \leftarrow \{\gamma^{(2)}\}$ &
$\{\{\alpha,\gamma^{(2)}\},\{\alpha^{(2)}\}\}$  $\{\{\alpha\},\{\alpha^{(2)}\},\{\gamma^{(2)}\}\}$  \\[3pt]
& $\{\{\alpha\},\{\alpha^{(2)},\gamma^{(2)}\}\}$ \\[3pt] \hline  & \\[-10pt]
$\{ \{\alpha^{(3)}\}\} \leftarrow \{\gamma^{(2)}\}$ & $\{ \{\alpha^{(3)}, \gamma^{(2)}\}\}$
$\{ \{\alpha^{(3)}\},\{\gamma^{(2)}\}\}$ \\[3pt] \hline
\end{tabular}
\\ \medskip{Table 4: Insertion of $\{\{\gamma^{(2)}\}\}$ in every submultiset
of $\{\alpha^{(3)}\}$}.
\end{center}

This strategy is speedier than the iterated full partition of
Andrews and Stafford \cite{Andrews1}, given that it takes into account the
multiplicity of all elements of $M.$ The higher this multiplicity is, the more
the insertion procedure gives efficient results, considering that
it involves more than one element of $M.$ At the end, we need to
compute the number of set partitions in $\Pi_5$ corresponding to
the same subdivision. For a given subdivision $S,$ this is
$$ \frac{\left( \begin{array}{c}
3 \\ \lambda_1, \lambda_2, \ldots, \lambda_k \end{array} \right)
\left( \begin{array}{c} 2 \\ \eta_1, \eta_2, \ldots, \eta_j
\end{array} \right)}{c_1! c_2! \cdots},$$ where $(\lambda_1,
\lambda_2, \ldots, \lambda_k)$ is the partition of $3$ giving the
number of times that $\alpha$ appears in every submultiset of $S,$
$(\eta_1, \eta_2, \ldots, \eta_j)$ is the partition of $2$ giving
the number of times that $\gamma$ appears in every submultiset of
$S$ and $c_t$ is the multiplicity of every submultiset in $S$.

As example for $\{\{\alpha, \gamma \},\{\alpha,
\gamma\},\{\alpha\}\}$ this number is $$ \frac{\left(
\begin{array}{c} 3 \\ 1, 1, 1 \end{array} \right) \left(
\begin{array}{c} 2 \\ 1, 1 \end{array} \right)}{1! 2!}.$$

When there are monomials involving correlated singleton umbrae in
the multiset $M$, see equivalence (\ref{eee}), we have further
speeded up the procedure. When in the same block of the
subdivision, there are monomials involving more than one
correlated singleton umbra, the evaluation of umbrae indexed by
this subdivision does not give contribution. Then, if we check the
indexes of singleton umbrae before the insertion  procedure,
we can delete the subdivision from the list and reduce the overall
computational time. For example, if we have $\{\{\chi_1 \alpha_1,
\chi_2 \alpha_2\}, \{\chi_2 \alpha_2\}\},$ the insertion of
$\chi_1 \alpha_1$ may be done only in the second set, because the
first one gives a zero contribution in the overall evaluation.

\subsection{$U$-statistics}
Table 5 shows computational times of three procedures implementing
the change of bases from augmented symmetric polynomials versus
power sums, which is at the bottom of the construction of
$U$-statistics. The three procedures are the function {\tt
AugToPower\-Sum} given in {\tt MathStatica} (release 1.0)
\cite{MathStatica}, the function {\tt TOP} given in {\tt SF}
(version 2.4) \cite{Stembridge}, and our {\tt MAPLE} function
{\tt augToPs}, with which we have implemented the second part of
(\ref{(ter3)}). Comparisons have shown how {\tt augToPs} performs
its task using less computational time than all the others.
\begin{center}
\begin{tabular}{|c||c||c||c|}
\hline & & &  \\[-10pt]
$[1^i 2^j 3^k \cdots]$  &  {\tt TOP}&  {\tt AugToPowerSum} & {\tt
augToPs} \\[3pt]
\hline & & & \\[-10pt]
$[1^5 \, 2^3 \, 3^2]$ & 0.78 & 0.18 & 0.13 \\[3pt]
\hline & & & \\[-10pt]
$[1^6 \, 2^3]$ & 0.08 & 0.01 & 0.01 \\[3pt]
\hline & & & \\[-10pt]
$[2^{10}]$ & 2.57 & 0.03 & 0.01 \\[3pt]
\hline & & & \\[-10pt]
$[1^5 \, 2^7 \, 3^1]$ & 6.15 & 1.20 & 0.65 \\[3pt]
\hline & & & \\[-10pt]
$[1^2 \, 2^2 \, 3^2 \,4^2]$ & 2.75 & 0.11 & 0.09 \\[3pt] \hline
\end{tabular}
\\ \medskip{Table 5: Comparison of computational times.}
\end{center}

Unlike our {\tt MAPLE} algorithm, note that {\tt AugToPowerSum}
and {\tt TOP} do not work on multiple sets of variables so no
comparisons can be done.

In order to compare the results achieved by means of umbral
methods with those of \cite{Andrews2}, we have performed
symbolic computations involved in unbiased estimators of product
of univariate and multivariate cumulants.

The symmetric statistic  $k_{r, \ldots,\, t}$ such that $E[k_{r,
\ldots,\, t}]=\kappa_r \cdots \kappa_t,$ where $\kappa_r, \ldots,
\kappa_t$ are univariate cumulants, is known as polykay. Being a
product of cumulants, the umbral expression of a polykay is simply
\begin{equation}
k_{r, \ldots,\, t} \simeq (\chi.\alpha)^r \cdots
(\chi^{\prime}.\alpha^{\prime})^t, \label{(pol1)}
\end{equation}
with $\chi, \ldots, \chi^{\prime}$ being uncorrelated singleton
umbrae and $\alpha, \ldots, \alpha^{\prime}$ satisfying $\alpha
\equiv \ldots \equiv \alpha^{\prime}.$ If $r + \cdots + t \leq n,$
the right-hand product of (\ref{(pol1)}) has the following umbral
expression  in terms of power sums:
\begin{equation}
k_{r,\ldots,t} = \sum_{(\lambda \, \vdash r, \, \ldots \,, \eta \,
\vdash t)} \!\!\! \frac{ (\chi.\chi)^{\nu_{\lambda}} \cdots
(\chi.\chi)^{\nu_{\eta}} \, d_{\lambda} \cdots
d_{\eta}}{(n)_{\nu_{\lambda} +  \cdots + \nu_{\eta}}} \!\!\!
\sum_{\pi \in \Pi_{\nu_{\lambda} +  \cdots + \nu_{\eta}}}
(\chi.\chi)^{.\pi} (n.\alpha)_{S_{\pi}}, \label{(polykaysfin)}
\end{equation}
where $S_{\pi}$ is the subdivision of the multiset
$$P_{\lambda+ \cdots + \eta} = \{\alpha^{(r_1 + \cdots +t_1)},
{\alpha^{2}}^{(r_2 + \cdots + t_2)}, \ldots \},$$ corresponding to
the partition $\pi \in \Pi_{\nu_{\lambda}+ \cdots + \nu_{\eta}}.$
In analogy with (\ref{(pol1)}), a multivariate polykay is a
product of multivariate cumulants, that is $E[k_{t_1 \ldots \,
t_r, \ldots, \, l_1 \ldots\, l_m}]= \kappa_{t_1 \cdots \, t_r}
\cdots \kappa_{l_1 \cdots \,l_m},$ where $\kappa_{t_1 \cdots \,
t_r}, \ldots, \kappa_{l_1 \cdots \,l_m}$ are multivariate
cumulants. Products of multivariate cumulants are represented by
products of uncorrelated multivariate $\alpha$-cumulant umbrae,
that is
\begin{equation}
k_{t_1 \ldots \, t_r, \ldots, \, l_1 \ldots\, l_m} \simeq
(\chi.\mu)_T \cdots (\chi^{\prime}.\mu^{\prime})_{L},
\label{(polmul)}
\end{equation}
where $\chi$ and $\chi^{'}$ are uncorrelated and the umbral
monomials $\mu \in T$ and $\mu^{\prime} \in L$ are such that
$$T=\{\mu_1^{(t_1)},\ldots, \mu_r^{(t_r)}\}, \, \ldots \,, L =
\{{\mu^{\prime}}_1^{(l_1)}, \ldots, {\mu^{\prime}}_m^{(l_m)}\}.$$
If $ |T|+ \cdots + |L| \leq n,$ the right-hand product of
(\ref{(polmul)}) has the following umbral expression in terms of
multivariate power sums:
\begin{equation}
k_{t_1 \ldots \, t_r, \ldots, \, l_1 \ldots\, l_m} \! = \!\!\!
\sum_{(\pi \in \Pi_{|T|}, \ldots, \tilde{\pi} \in \Pi_{|L|})}
 \!\!\! \!\!\! \frac{(\chi.\chi)^{|\pi|} \cdots
(\chi^{\prime}.\chi^{\prime})^{|\tilde{\pi}|}} {(n)_{|\pi| +
\cdots + |\tilde{\pi}|}} \!\!\!\!\!\! \sum_{\tau \in \Pi_{|\pi| +
\cdots + |\tilde{\pi}|}} \!\!\! (\chi.\chi)^{.\tau}
(n.p)_{S_\tau}, \label{(polmul1)}
\end{equation}
where $S_{\tau}$ is the subdivision of the multiset obtained by
the disjoint union of $T, \ldots, L$ with no uncorrelation labels
and corresponding to the partition $\tau$ of the set built with
the blocks of $\{\pi,\ldots,\tilde{\pi}\}.$ The umbral formulae
here recalled are stated in \cite{DiNardo3}.

Table 6 shows computational times obtained by the procedure {\tt
PolyK} of {\tt Math\-Statica}\footnote{Of course,
if single $k$-statistics are enough to be computed, the function {\tt k-stat}
of {\tt MathStatica} is suitable and faster.},  by our {\tt MAPLE} function {\tt
polyk} implementing (\ref{(polmul)}), and by the procedures
proposed by Andrews and Stafford \cite{Andrews1} (in particular, see \cite{Bellhouse}
for multiple sums). Note that (\ref{(polmul1)}) gives as special
case both (\ref{(pol1)}) and univariate and multivariate
$k$-statistics (\ref{(dotprod12)}). We remark that {\tt
MathStatica} has no procedure to handle multivariate polykays
\footnote{In the forthcoming {\tt MathStatica}, release 2,
the procedure to handle multivariate polykays is now availbale
(C. Rose, private communication).} The
computational times of  Andrews and Stafford's procedures have
been obtained by the code available at
http://fisher.utstat.toronto.edu/david/SCSI/chap.3.nb.
\begin{table}[h]
\begin{center}
\begin{tabular}{|c||c||c||c|} \hline\\[-10pt]
$k_{t, \ldots,\, l}$ & Andrews-Stafford & {\tt MathStatica} & {\tt
MAPLE} \\ \hline & & \\[-11pt]
$k_{8}$              &     0.10    & 0.09   &  0.05 \\[3pt]
\hline & & \\[-11pt]
$k_{10}$             &     0.35    & 0.25   &  0.15 \\[3pt]
\hline & & \\[-11pt]
$k_{12}$             &     1.19    & 0.84   &  0.42 \\[3pt]
\hline & & \\[-11pt]
$k_{14}$             &     3.93    & 2.67   &  1.29 \\[3pt]
\hline & & \\[-11pt]
$k_{16}$             &    12.74    & 8.54   &  3.86 \\[3pt]
\hline & & \\[-11pt]
$k_{18}$             &    40.44    & 30.32  &  12.3 \\[3pt]
\hline & & \\[-11pt]
$k_{6,6}$            &    39.84    &  0.81  &  0.51 \\[3pt]
\hline & & \\[-11pt]
$k_{9,3}$            &    18.32    &  0.84  &  0.46 \\[3pt]
\hline & & \\[-11pt]
$k_{9,6}$            &    676.14   &  5.02  &  3.13 \\[3pt]
\hline & & \\[-11pt]
$k_{9,9}$            &   $>1.5$hh  & 29.10  & 23.19 \\[3pt]
\hline & & \\[-11pt]
$k_{3,3} k_{2,2}$    &     4.67    & \hbox{n.c.}   &        1.71 \\[3pt]
\hline & & \\[-11pt]
$k_{3,3} k_{3,3}$    &    32.16    & \hbox{n.c.}   &       15.87 \\[3pt]
\hline & & \\[-11pt]
$k_{2,1,1} k_{2,1,1}$&     1.031   & \hbox{n.c.}   &        0.52 \\[3pt]
\hline
\end{tabular}
\\ \medskip{Table 6: Comparison of computational times. \\ The acronym \lq\lq n.c.\rq\rq stands
for not calculable.}
\end{center}
\end{table}

Remark that the output expressions of Andrews and Stafford's code
are unpractical. These are very different from the output
expressions of {\tt PolyK} of {\tt MathStatica} and {\tt polyk} in
{\tt MAPLE}. For example, the output of Andrews and Stafford's
code for $k_3$ is
\begin{equation}
\scriptstyle{\frac{2
\bar{X}^3}{\left(1-\frac{2}{n}\right)\left(1-\frac{1}{n}\right)} +
\left( - \frac{3}{1-\frac{1}{n}} -
\frac{6}{\left(1-\frac{2}{n}\right)\left(1-\frac{1}{n}\right)n}
\right) \bar{X} \, \overline{X^2} + \left( 1 +
\frac{4}{\left(1-\frac{2}{n}\right)\left(1-\frac{1}{n}\right)n^2}
+ \frac{3}{\left(1-\frac{1}{n}\right)n} \right) \overline{X^3}},
\label{(ex)}
\end{equation}
whereas the output of {\tt PolyK} of {\tt MathStatica} and {\tt
polyk} in {\tt MAPLE} is
$$\frac{n^2 S_3 - 3 n S_1 S_2 + 2 S_1^3}{n(n-1)(n-2)}.$$
In $k_{12},$ the expression  of Andrews and Stafford's code
consists of $602$ terms compared with $77$ terms of the
expressions obtained by {\tt PolyK} of {\tt MathStatica} and {\tt
polyk} in {\tt MAPLE}. In order to recover the same output of {\tt
PolyK} of {\tt MathStati\-ca} and {\tt polyk} in {\tt MAPLE} in
(\ref{(ex)}), we must group the terms in parenthesis over a common
denominator, deleting equal factors in the results. This operation
increases the overall computational time. For example the
computational time of $k_{10}$ grows from $0.35$ to $2.693,$ the
one of $k_{12}$ grows from $1.191$ to $14.56.$ For $k$-statistics
of order greater than $14,$ an error occurs since the recursion
exceeds a depth of $256.$
\subsection{Product of augmented symmetric functions}
The {\tt MAPLE} routine {\tt Pam} implements equivalences such as
(\ref{eee}). {\tt Pam} calls the routine {\tt makeTab}.
Considering that there are monomials involving singleton umbrae in
the multiset $M,$ the efficiency of {\tt makeTab} improves, as we
have mentioned at the end of Subsection 6.1. In Table 7, we
compare some computational times of {\tt Pam} with those of the
routine {\tt SIP}, written in {\tt MATHEMATICA} language by Vrbik
\cite{Vrbik}, and exclusively devoted to products of augmented
symmetric functions.
\begin{center}
\begin{tabular}{|c||c||c|} \hline \\[-10pt]
\strut $[1^i 2^j 3^k \cdots]$ &  {\tt SIP} &  {\tt MAPLE}  \\[1pt]
\hline & & \\[-10pt]
\strut $[5^3 \, 9 \, 10][1 \, 2 \, 3 \, 4 \, 5]$             &    0.7     &      0.1 \\[1pt]
\hline & & \\[-10pt]
\strut $[5^3 \, 8 \, 9 \, 10][1 \, 2 \, 3 \, 4 \, 5]$           &    5.6     &      0.4 \\[1pt]
\hline & & \\[-10pt]
\strut $[6 \, 7 \, 8 \, 9 \, 10][1 \, 2 \, 3 \, 4 \, 5]$           &    2.2     &      0.1 \\[1pt]
\hline & & \\[-10pt]
\strut $[6 \, 7 \, 8 \, 9 \, 10][1 \, 2][3 \, 4 \, 5]$       &    3.1     &      0.4 \\[1pt]
\hline & & \\[-10pt]
\strut $[6 \, 7][8 \, 9 \, 10][1 \, 2][3 \, 4 \, 5]$      &    4.7     &      1.3 \\[1pt]
\hline & & \\[-10pt]
\strut $[5 \, 6 \, 7 \, 8 \, 9 \, 10][1 \, 2 \, 3 \, 4 \, 5]$         &   16.7     &      0.3 \\[1pt]
\hline & & \\[-10pt]
\strut $[5 \, 6 \, 7 \, 8 \, 9 \, 10][1 \, 2 \, 3 \, 4 \, 5 \, 6]$       &  348.7     &      1.5 \\[1pt]
\hline & & \\[-10pt]
\strut $[6 \, 7 \, 8 \, 9 \, 10][6\, 7][3 \, 4 \, 5][1\, 2]$   & 125.6     &     16.4 \\[1pt]   \hline
\end{tabular}
\\ \medskip{Table 7: Comparison of computational times.}
\end{center}

Note that the computational time of {\tt SIP} depends heavily on
the number of variables involved in the brackets, because it
resorts set permutations, whose computation cost is factorial in
its cardinality.
\section{Concluding remarks}
This paper focuses attention on a symbolic calculation of products
of statistics related to cumulants or moments. Undoubtedly, an
enjoyable challenge is to find efficient procedures to deal with
the necessarily huge amount of algebraic and symbolic computations
involved in such a kind of calculations. The methods we propose
result more efficient compared with those available in the
literature. Note that high order statistics have a variety of
applications. Recently, Rao \cite{Rao} have shown applications of high
order cumulants in statistical inference and time series. Indeed,
there are different areas, such as astronomy (see \cite{Prasad}
and references therein), astrophysics \cite{Ferreira} and
biophysics \cite{Muller}, where one computes high order
$k$-statistics in order to recognize a gaussian population or
characterizes asymptotic behavior of high order $k$-statistics if
the population is gaussian. Indeed, $k$-statistics are independent
from the sample mean if and only if the population is gaussian
\cite{Lukas} and in such a case $k$-statistics of order greater
than $2$ should be nearly to zero. For such applications,
increasing speed and efficiency is a significant investment.

As we have shown, the codes of Andrews and Stafford are quite
inefficient for the problems posed here. This paper has pointed
out the role played by the notion of subdivision in speeding up
the calculations resulting by multiplying sums of random variables
and the role played by the umbra $\chi$ in selecting the involved
variables. The symbolic algorithm we propose, in order to
evaluating the mean of product of augmented polynomials in random
variables, relies on this innovative strategy.

In closing, we would like to emphasize that classical umbral
calculus not only decreases the computational time, but offers a
theory to prove more general results. Recently, L-moments and
trimmed L-moments have been noticed as appealing alternatives to
 conventional moments, see \cite{Karvanen} and
\cite{Delicado}. We believe that the handling of these number
sequences would benefit by an umbral approach.

\section{Appendix}

In the following, we present the MAPLE code of the
procedure giving subdivisions of a multiset. In order to have the
results of Table 3 and 4, the calling syntax is {\tt
makeTab(3,2)}.

 {\footnotesize
\begin{verbatim}
nRep := proc(u)
  mul(x[2]!,x=convert(u,multiset)); end:
#- - - - - - - - - - - - - - - - - - - - - - - - - - - - -#
URv := proc(u,v)
local U,ou,i,ptr_i,vI;
ou:=NULL; U:=[]; vI:=indets(v);
for ptr_i from nops(u) by -1 to 2 do
    if has(u[ptr_i],v) then break; fi; od;
for i from ptr_i to nops(u) do
    if not (u[i]=ou or has(u[i],vI)) then ou:=u[i];
      U:=[op(U),[op(u[1..(i-1)]),u[i]*v,op(u[(i+1)..-1])]];
    fi;
od; op(U),[op(u),v]; end:
#- - - - - - - - - - - - - - - - - - - - - - - - - - - - -#
URV := proc()
local U,V,i;
U:=[args[1,1]]; V:=args[2,1];
for i from 1 to nops(V) do
    U := [ seq( URv(u,V[i]), u=U ) ];
od; seq([x,args[1,2]*args[2,2]/nRep(x)],x=U) ; end:
#- - - - - - - - - - - - - - - - - - - - - - - - - - - - -#
URmV := proc()
  local U,i,nbin;
  if nargs=1 then
       U:=args;
  else U:=URV( args[1], args[2]);
       for i from 3 to nargs do
           U:=seq( URV( u, args[i]), u=[U] );
       od; fi;
  U; end:
#- - - - - - - - - - - - - - - - - - - - - - - - - - - - -#
comb := proc(V,ptr,Y)
 if ptr=nops(V)+1 then return(Y); fi;
 seq( comb(V, ptr+1, [ op(Y), L ] ), L=V[ptr]); end:
#- - - - - - - - - - - - - - - - - - - - - - - - - - - - -#
makeTab := proc()
local U;
U:=[seq( [seq( [[seq(P||i^z,z=y)],
              combinat['multinomial'](args[i],seq(r,r=y))],
      y=combinat['partition'](args[i]))],
    i=1..nargs)];
if nops(U)=1 then [seq([x[1],x[2]/nRep(x[1])],x=op(U))];
             else [seq(URmV(op(x)),x=[comb(U,1,[])])];
fi; end:
\end{verbatim}
}

\section{Acknowledgments}
The authors thank the referees for valuable comments and
suggestions, which improved the presentation of the paper.
Moreover, the authors thank Pasquale Petrullo for his contribution
in optimizing the algorithm {\tt makeTab}.

\end{document}